\documentclass[a4paper,fleqn]{cas-sc} 

\usepackage[numbers]{natbib}

\usepackage{graphicx}
\usepackage{enumitem}
\usepackage{caption}
\usepackage{hyperref}
\usepackage{multirow}
\usepackage{amsmath}
\usepackage{amssymb}
\usepackage{rotating}
\usepackage{caption}
\usepackage{subcaption}
\usepackage{booktabs}
\usepackage{float}
\usepackage{pdflscape}

\newcommand{\partialdiff}[2] {\dfrac{\partial {#1}} {\partial {#2}}}

\newcommand{\partialdiffs}[2] {\frac{\partial {#1}} {\partial {#2}}}

\newcommand{\brac}[1]  {\left( #1 \right)}

\newcommand{\bigbrac}[1]  {\left\{ #1 \right\}}

\newcommand{\norm}[1]  {\left\| #1 \right\|}
\newcommand{\dummy}  {\{~\}}

\newcommand{\divergence}[1]{\nabla \cdot #1 }

\newcommand{\laplace}[1]{\nabla^2 #1}

\newcommand{\VecU}{\mathbf{U}}
\newcommand{\model}[1]{\ensuremath{\mathcal{M}_{\textrm{#1}}}}

\def\tsc#1{\csdef{#1}{\textsc{\lowercase{#1}}\xspace}}
\tsc{WGM}
\tsc{QE}

\begin{document}
\let\WriteBookmarks\relax
\def\floatpagepagefraction{1}
\def\textpagefraction{.001}

\shorttitle{Prediction of Steady-State Flow through Porous Media Using Machine Learning Models}    

\shortauthors{J. Wang et~al.}  

\title [mode = title]{Prediction of Steady-State Flow through Porous Media Using Machine Learning Models}  
    


%

\author[1]{Jinhong Wang}[orcid=0000-0002-0012-7218,]
\fnmark[1]




\affiliation[1]{organization={Department of Mechanical Engineering, Imperial College London},
            addressline={South Kensington Campus}, 
            city={London},
            postcode={SW7 2AZ}, 
            state={},
            country={United Kingdom}}

\author[1]{Matei C. Ignuta-Ciuncanu}
\fnmark[1]

\author[1]{Ricardo F. Martinez-Botas}[orcid=0000-0002-3263-8151]

\author[1]{Teng Cao}[orcid=0009-0000-2594-9157]
\cormark[1]
\ead{t.cao@imperial.ac.uk}
\cortext[1]{Corresponding author}

\begin{abstract}
    Solving flow through porous media is a crucial step in the topology optimisation of cold plates, a key component in modern thermal management. Traditional computational fluid dynamics (CFD) methods, while accurate, are often prohibitively expensive for large and complex geometries. In contrast, data-driven surrogate models provide a computationally efficient alternative, enabling rapid and reliable predictions.
    In this study, we develop a machine-learning framework for predicting steady-state flow through porous media governed by the Navier-Stokes-Brinkman equations. We implement and compare three model architectures—convolutional autoencoder (AE), U-Net, and Fourier Neural Operator (FNO)—evaluating their predictive performance. To enhance physics consistency, we incorporate physics-informed loss functions. Our results demonstrate that FNO outperforms AE and U-Net, achieving a mean squared error (MSE) as low as 0.0017 while providing speedups of up to 1000× compared to CFD. Additionally, the mesh-invariant property of FNO emphasizes its suitability for topology optimisation tasks, where varying mesh resolutions are required.
    This study highlights the potential of machine learning to accelerate fluid flow predictions in porous media, offering a scalable alternative to traditional numerical methods. 
\end{abstract}





\maketitle

\section{Introduction}

Effective thermal management is critical in modern engineering, ensuring the reliability and efficiency of systems ranging from electronics to energy applications. 
As devices have become more compact, the need for efficient heat dissipation has led to the development of cold plates, which transfer heat from high-heat-flux sources to cooling mediums. 
To improve upon this, heat pipes were introduced, integrating smaller, more efficient internal features for passive heat transfer. 
Building on these advancements, the optimization of flow architectures has led to the creation of designed porous media \cite{Bejan2004}, where geometry is free to adapt for maximum performance, enhancing heat transfer by controlling fluid flow and increasing surface area for more targeted thermal solutions.

Among the emerging thermal management technologies, both cold plates \cite{mo2021topology} and heat pipes \cite{Lurie2019} rely on precise hydrodynamic models to optimize their performance. 
Density-based topology optimization is often employed to enhance thermal performance and minimize pressure losses \cite{Mo2021, boutsikakis2024, Wu2024} by tailoring flow distribution and material placement, using sensitivity analysis to compute the gradients of the objective function with respect to the design (density) variable.
Alternatively, generative constructal design approaches allow configurations to emerge as a result of evolutionary optimization, where optimal flow paths and geometries evolve to meet thermal performance objectives \cite{ATP2023}.
Supervised learning methods have been successfully applied to design conductive heat sinks \cite{ATP2023} and macroscopic thermal metamaterials \cite{IgnutaTabor2024}. 
On the other hand, unsupervised learning has been used to identify non-uniform dendritic structures that effectively meet hierarchical design objectives \cite{CLC2024}.

Regardless of the design method, a critical step in this optimization process is the accurate calculation of the heat and/or fluid flow physics.
Within a modelled porous medium, this is notoriously difficult due to the complex interactions between the fluid and the medium's geometry, which can vary significantly across the system.
Topology optimization retrieves the flow field through a modelled porous media by numerically solving the modified Navier–Stokes equations with a Brinkman term representing the fluid interactions with a moving boundary \cite{mo2021topology}.
This calculation is computationally expensive and time-consuming, especially for complex geometries. 
The general formulation enables the development of machine learning surrogate models for broader classes of flows in porous media, effectively capturing complex flow patterns and accommodating varying levels of material anisotropy, extending applications beyond Darcy flow \cite{li2020fourier}. 

With the fast-developing computing power, machine learning methods have become a promising alternative to traditional numerical methods when solving partial differential equations \cite{LeCun2015, Brunton2020}. \citet{raissi2019physics} proposed physics-informed neural networks (PINN) that embed physics laws in the learning process, offering a robust framework for solving complex, high-dimensional PDEs while enforcing physical consistency. While many neural network architectures developed in the literature have been tested on simplified benchmark examples, there is a growing interest in applying such models to real-world engineering challenges, enabling rapid and reliable solutions for complex systems like aerodynamics \cite{li2022machine}, energy systems \cite{mosavi2019state}, and thermal management \cite{al2023batteries}.

\citet{wang2024stacked} introduced a weakly supervised machine learning approach for solving steady-state Navier–Stokes equations, using a physics-informed loss and multichannel inputs for boundary and geometric conditions. 
\citet{Ruilong} proposed a grid-free deep learning method based on a physics-informed neural network to solve coupled Stokes–Darcy equations, for a fixed boundary between free and porous flow.
\citet{Niekamp2023} proposed a data-driven surrogate model for predicting permeabilities and laminar flow in random micro-heterogeneous materials, using a stochastic model to generate training data and comparing UResNet and FCNN architectures.
However, these studies have primarily focused on the Darcy flow instead of the more general Navier-Stokes-Brinkman that can model moving boundaries. 
In addition, the literature usually only focuses on single model architecture without performance behaviours between different models.

In this work, we propose a machine-learning framework for predicting steady-state flow through porous media, modeled by the Navier-Stokes-Brinkman equations which directly addresses viscous resistance effects inherent in porous flow problems. 
We evaluate three architectures: autoencoders \cite{bank2023autoencoders}, U-Net \cite{ronneberger2015u}, and Fourier Neural Operator (FNO) \cite{li2020fourier}, highlighting their strengths and trade-offs in terms of accuracy and computational efficiency. 
Physics-informed losses are introduced and compared with the baseline models to show the model performance enhancement.
This study demonstrates the potential of neural networks to accelerate and generalize fluid flow predictions in porous systems, offering a scalable alternative to traditional numerical methods. 
\section{Methodology}

To address the numerical challenges and high computational cost associated with solving the Brinkman equation, we explore a machine learning-based approach to accelerate the solution process of flow through porous media. As a proof of concept, we consider a 2D steady-state incompressible flow of a Newtonian fluid through a porous medium. Specifically, we test and compare three machine learning architectures: convolutional autoencoder, U-Net, and Fourier Neural Operator (FNO). These models are trained on solution data using finite element method (FEM) to learn efficient representations of the solution space, enabling faster and more scalable predictions. To enhance accuracy and physical consistency, we incorporate a physics-informed loss function that enforces key governing principles of the Brinkman equation. This hybrid data-driven and physics-informed framework aims to provide a robust and computationally efficient alternative to traditional numerical solvers.

\subsection{Physical Problem Setup}

\subsubsection{Case Setup}
\begin{figure}
    \centering
    \includegraphics[width=\linewidth]{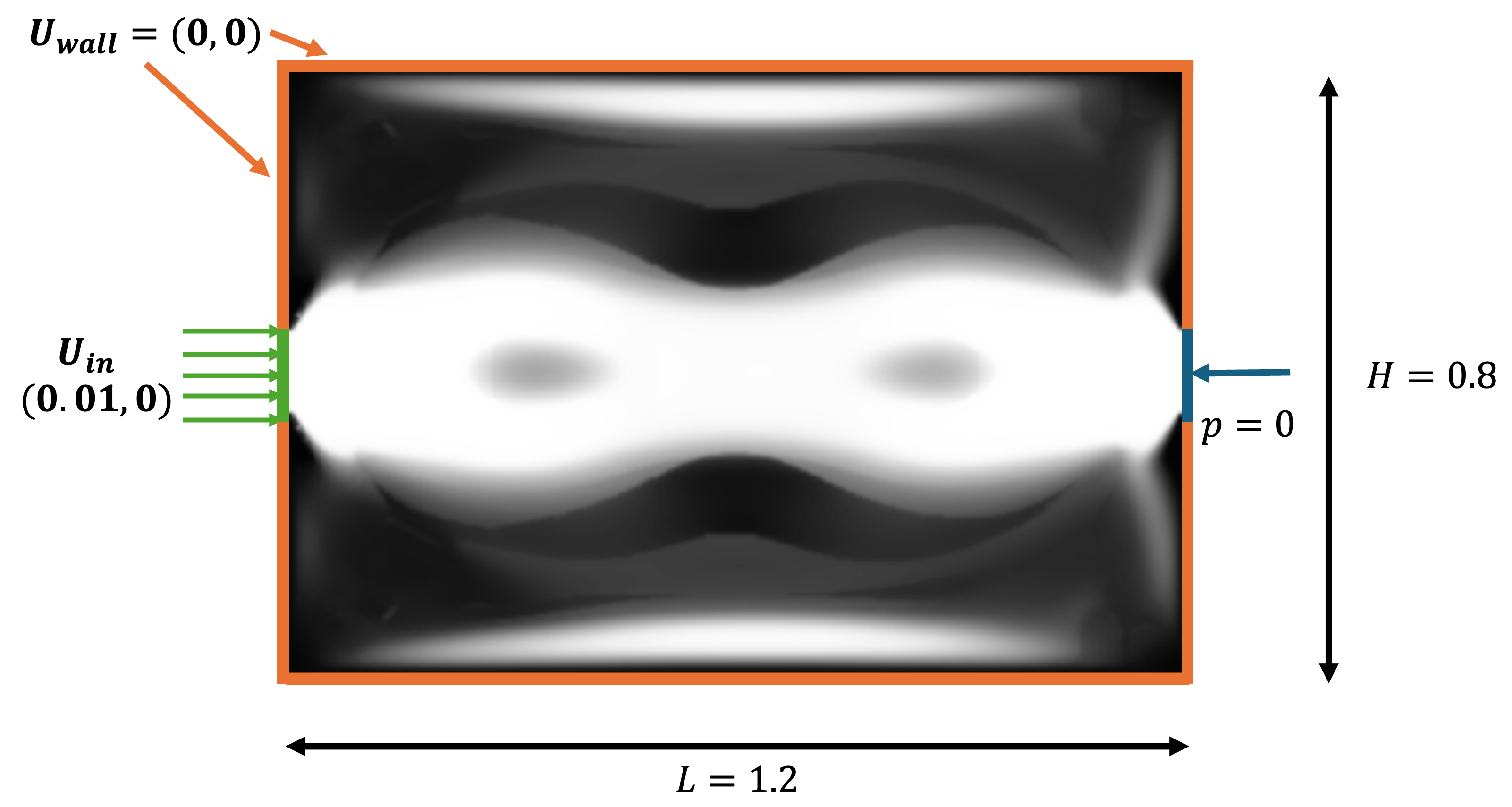}
    \caption{Computational domain for the finite element analysis (FEM), showing the boundary conditions and design parameters used in the optimization process.}
    \label{fig:domain}
\end{figure}

Figure~\ref{fig:domain} presents an illustration of the physics domain. The fluid density $\rho$ and dynamic viscosity $\mu$ are assumed to be constant, and the flow is governed by the Brinkman equation (Eq.~\eqref{brinkman equation}) to account for both inertial and viscous effects.

The boundary conditions for this problem are summarised in Table~\ref{tab:BCs}. The inlet velocity, \( U_{\text{in}} \), is specified as \( 0.01 \, \text{m/s} \), corresponding to a Reynolds number of \( Re = 100 \), based on the characteristic length and kinematic viscosity of the fluid. 
A zero-gauge pressure condition is applied at the outlet boundary and a no-slip boundary condition is imposed on all solid walls.

\begin{table}[width=\linewidth, cols=2, pos=ht]
    \setlength{\tabcolsep}{6pt}
    \renewcommand{\arraystretch}{1.5}
    \centering
    \caption{Boundary conditions used in the simulation}

    \begin{tabular*}{\tblwidth}{@{} ll@{} }
        \toprule
        \textbf{Boundary Condition} & \textbf{Equation} \\
        \midrule
        \textbf{Inlet Velocity} & \( \VecU = \left( 0.01, 0 \right) \quad \text{on the inlet boundary} \) \\
        \textbf{Outlet Pressure} & \( P = 0 \quad \text{on the outlet boundary} \) \\
        \textbf{Wall (No-Slip)} & \( \VecU = (0, 0) \quad \text{on the wall boundaries} \) \\
        \bottomrule
    \end{tabular*}
    \label{tab:BCs}
\end{table}

\subsubsection{Governing Equations and Physical Constraints}

Fluid flow through porous media is commonly modelled within a continuum framework, as detailed in \cite{NieldBejan2013}. 
In this approach, the Darcy velocity is defined as the average velocity over a representative elementary volume (REV) that encompasses both solid and fluid phases. 
This formulation leads to the standard expression for mass conservation, represented by the continuity equation.

The conservation of momentum in porous media can be described by different models, each applicable under specific conditions. 
One widely used model is Darcy's Law, 
\begin{equation}
    \label{darcy's equation}
    \nabla P = - \frac{\mu}{K} \VecU
\end{equation}
which relates the seepage velocity of the fluid ($\VecU$) of the fluid to the pressure gradient and the permeability ($K$) of the porous medium. 
The permeability, a property of the medium, quantifies the ease with which fluid flows through its pores.

For cases where inertial effects become significant, the work of Wooding \cite{Wooding1957} was followed to provide an extended framework that includes inertial terms in the momentum equation.
\begin{equation}
    \label{ns equation}
     \rho \brac{ \partialdiff{\VecU}{t} + \VecU \cdot \nabla \VecU} = - \nabla P - \frac{\mu}{K} \VecU
\end{equation}

When addressing scenarios where the flow occurs near solid boundaries as well as through clear fluid, the Brinkman equation as Eq.~\eqref{brinkman equation} can be employed, as the permeability can be used to switch between the Navier-Stokes ($K \to \infty$) and Darcy ($K \to 0$) regimes \cite{NieldBejan2013}.
\begin{equation}
    \label{brinkman equation}
     \rho \brac{ \partialdiff{\VecU}{t} + \VecU \cdot \nabla \VecU} = - \nabla P + \Tilde{\mu} \laplace{\VecU} - \alpha \VecU
\end{equation}
This equation adds a Laplacian drag term ($\laplace{\VecU}$) in analogy with the Navier-Stokes equation, and lumps the effective viscosity $\mu/K$ into a geometry dependent parameter $\alpha$.
The resulting Navier-Stokes-Brinkman model enables the modelling of velocity gradients and boundary layer effects in porous materials \cite{NieldBejan2013}.
As the most general formulation for porous flow, we adopt the Brinkman equation in this study to ensure a comprehensive coverage of a wide range of flow conditions.
In addition, this methodology can be extended to applications such as cold-plate topology optimization and natural convection during phase change in the presence of solid obstacles, as demonstrated in \cite{SHTC25}.

For data collection, the system of equations governing fluid flow in porous media (Eqs .~\eqref{incompressible_ns}) is solved using the finite element method (FEM), implemented in the open-source FEniCS library \cite{LoggWells2010}.

\begin{equation}
    \left\{
    \begin{aligned}
        \divergence \VecU &= 0, \\
        \rho \brac{ \partialdiff{\VecU}{t} + \VecU \cdot \nabla \VecU} &= - \nabla P + \Tilde{\mu} \laplace{\VecU} - \alpha \VecU.
    \end{aligned}
    \right.
    \label{incompressible_ns}
\end{equation}

The domain is defined as a rectangular region and discretized into a mesh of mixed finite elements to ensure compatibility between velocity and pressure fields \cite{Logg2012}.
Quadratic Lagrange elements are used for velocity, while linear Lagrange elements are used for pressure. 
The governing equations (Eq.~\eqref{incompressible_ns}) are reformulated into their finite element weak forms. 

The permeability field is defined using the design variable $\gamma$, allowing for spatial variation of porous properties.
The drag term proportional to the velocity introduced by the Brinkman term ($- \alpha \VecU$) in Eq.~\eqref{ns equation} is parameterized by a permeability-dependent coefficient $\alpha(\gamma)$. 
The interpolation for $\alpha(\gamma)$ is defined using the RAMP (Rational Approximation of Material Properties, \cite{mo2021topology}) method:
\begin{equation}
    \alpha(\gamma) = \alpha_{\text{min}} + (\alpha_{\text{max}} - \alpha_{\text{min}}) \frac{1 - \gamma}{1 + q_a \gamma},
\end{equation}
where \(\alpha_{\text{min}}\) and \(\alpha_{\text{max}}\) are the minimum and maximum drag coefficients, respectively, and \(q_a\) is a penalization parameter controlling the transition sharpness. This formulation ensures a smooth variation of \(\alpha(\gamma)\) between highly permeable and nearly impermeable regions, enhancing numerical stability and physical realism.

The weak form leads to a nonlinear system, primarily due to the convective terms in the momentum equation and the coupling between velocity and pressure. 
This system is assembled within the finite element framework and solved using a Newton-based approach. 
The FEniCS backend is configured with PETSc's Scalable Nonlinear Equations Solvers (SNES), leveraging the MUltifrontal Massively Parallel Sparse (MUMPS) direct solver for the linearized equations and the hypre-AMG preconditioner.
The system of equations is solved iteratively for the velocity ($\VecU$) and pressure ($p$) fields.
Convergence is achieved when the residuals of the equations meet predefined tolerances of $10^{-6}$ for both absolute and relative errors.

The topology optimization algorithm samples 100 different parameters from a normal distribution with specified means and standard deviations for each parameter, including $p_{SIMP}$, $\alpha_{max}$, $q_a$ and $w$.
It then runs for 30 iterations using the Method of Moving Asymptotes (MMA) to iteratively update the design variable $\gamma$ in a multi-objective thermal optimization process for cold plates.
The optimized values of the design variable ($\gamma$), and the corresponding velocity ($\VecU$), and pressure fields ($p$) are interpolated to the vertices of the rectangular mesh and saved as flat arrays.

\begin{table}[width=.6\linewidth,cols=2,pos=ht]
    \setlength{\tabcolsep}{6pt}
    \renewcommand{\arraystretch}{1.5}
    \centering
    \caption{Cold Plate Design Parameters}
    \label{tab:parameters}
    \begin{tabular*}{\tblwidth}{@{}CC@{}}
        \toprule
        \textbf{Parameter} & \textbf{Distribution} \\ \midrule
        \( p_{SIMP} \) & $\mathcal{N}(0.01, 0.001)$ \\
        \( \alpha_{\text{max}} \) &  $\mathcal{N}(10000, 1000)$ \\
        \( q_a \) &  $\mathcal{N}(100, 10)$ \\
        \( w \) &  $\mathcal{N}(0.5, 0.1)$ \\
        \bottomrule
    \end{tabular*}
\end{table}

\subsection{Machine Learning Model Architecture}
\label{Model Architecture method section}

The neural network models \model{} in this paper are represented as a transformation $\model{}: \mathcal{A} \rightarrow \mathcal{U}$ where $\mathcal{A}$ and $\mathcal{U}$ denote the input and output spaces, respectively. 
The models take as input a single-channel 2D field $\alpha\brac{x,y}$, representing the material's porosity distribution, and output a three-channel field $\widehat{\Phi}\brac{x,y}$, which contains the predicted flow properties as described in Eq.~\eqref{model general equation}.
\begin{equation}
    \label{model general equation}
    \begin{split}
        \widehat{\Phi} \brac{x,y} &= \bigbrac{ \widehat{P} \brac{x,y}, \widehat{U}_x \brac{x,y}, \widehat{U}_y \brac{x,y}} \\
        &= \model{} \Big( \alpha\brac{x,y} \Big) \\
        &~~\text{where}~~~ \alpha \in \mathcal{A} ~,~ \widehat{\Phi} \in \mathcal{U} 
    \end{split}
\end{equation}
The output $\widehat{\Phi}\brac{x,y}$ includes the predicted pressure field $\widehat{P}$ and the velocity components $\widehat{U}_x$ and $\widehat{U}_y$ in $x$- and $y$-directions, respectively.





\begin{figure*}
    \centering
    \includegraphics[width=\textwidth]{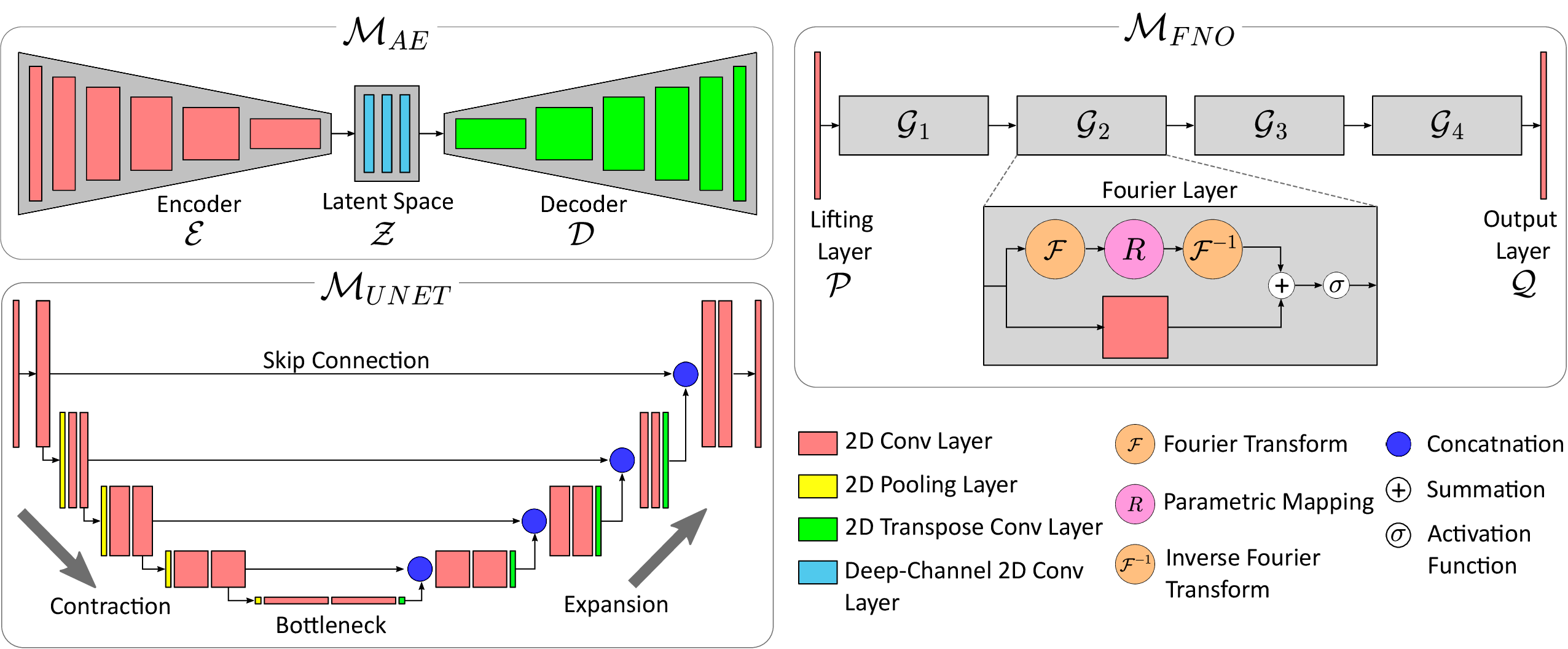}
    \caption{Model architectures illustrations for \model{AE}, \model{UNET}, and \model{FNO}}
    \label{model architecture illustration}
\end{figure*}

Three models are trained and compared in this paper: convolutional autoencoder (AE) \cite{bank2023autoencoders}, U-net \cite{ronneberger2015u}, and the Fourier Neural Operator (FNO)\cite{li2020fourier}. An illustration of the model architectures used in this paper is summarised in Fig.~\ref{model architecture illustration}. To provide fairer comparisons between the model performance, all three architectures are adjusted to have a similar amount of parameters as listed in Table~\ref{Model Parameter Information}.

\begin{table}[width=\linewidth,cols=3,pos=ht]
    \setlength{\tabcolsep}{6pt}
    \renewcommand{\arraystretch}{1.5}
    \centering
    \caption{Model Parameter Information}

    \begin{tabular*}{\tblwidth}{@{} CCC@{} } 
        
        \toprule
        Model Name &  Symbol & Parameter Number \\ 
        \midrule
        
        Convolutional Autoencoder & \model{AE} & 30389123 \\

        U-Net & \model{UNET} & 31042499 \\ 

        Fourier Neural Operator & \model{FNO} & 29516675 \\ 
        
        \bottomrule
        
    \end{tabular*}

    \label{Model Parameter Information}
\end{table}


\subsubsection{Convolutional Autoencoder \model{AE}}
\label{ae architecture section}

A symmetric convolutional autoencoder with deep-channel latent layers is developed, denoted as \model{AE}. The model is formulated as:
\begin{equation}
    \model{AE} \brac{\alpha} = \mathcal{D} \circ \mathcal{Z} \circ \mathcal{E} \brac{\alpha}
\end{equation}
where $\mathcal{E}$, $\mathcal{Z}$, and $\mathcal{D}$ represent the encoder, latent space, and decoder layers, respectively. As shown in Fig.~\ref{model architecture illustration}, the encoder $\mathcal{E}$ contains six 2D convolutional layers, each with ReLU activation, a kernel size of $3 \times 3$, and a stride of $2$. The latent space layers $\mathcal{Z}$ contain three deep-channel 2D $3 \times 1$ convolutional layers with ReLU activation and a kernel size of 1. The decoder $\mathcal{D}$ comprises six transposed convolutional layers to reconstruct the flow field from the latent space features, using the same activation function, kernel size, and stride as $\mathcal{E}$.

\subsubsection{U-NET \model{UNET}}
\label{unet architecture section}

The U-net model with a depth of 4 is introduced in this paper as shown in Fig.~\ref{model architecture illustration}. Similar to \model{AE}, U-net architecture consists of an encoder-decoder (contraction-expansion) structure but with additional skip connections, enabling the network to capture both high-level contextual and low-level detailed features. The encoder path consists of a series of convolutional blocks, each followed by a max-pooling layer that downsamples the input by a factor of 2 (stride of 2). The convolutional blocks each contain two $3 \times 3$ convolutional layers with padding to preserve the spatial dimensions, followed by ReLU activations. The number of channels doubles at each level of the encoder, starting from an initial number of 64 filters, to capture increasingly complex features.

At the bottleneck of the network, a final convolutional block processes the deepest features before the decoder path. The decoder path consists of upsampling operations using $3 \times 3$ transposed convolutions (stride of 2), which double the spatial dimensions, followed by concatenation with the corresponding feature maps from the encoder through skip connections. Each decoder block contains two $3 \times 3$ convolutions followed by ReLU activations. Finally, an output convolutional layer maps the output to the desired number of channels.

\subsubsection{Fourier Neural Operator \model{FNO}}
\label{FNO architecture section}

As shown by Fig.~\ref{model architecture illustration}, \model{FNO} in this paper follows the base model architecture presented in \cite{li2024physics}, incorporating four Fourier layers $\mathcal{G}$ positioned between the lifting layer $\mathcal{P}$ and the output layer $\mathcal{Q}$: 
\begin{equation} 
    \model{FNO} \brac{\alpha} = \mathcal{Q} \circ \mathcal{G}_4 \circ \mathcal{G}_3 \circ \mathcal{G}_2 \circ \mathcal{G}_1 \circ \mathcal{P} \brac{\alpha}. 
\end{equation} 
The lifting layer $\mathcal{P}$ uses a convolutional layer with $1 \times 1$ kernel to project the input $\alpha$ into a tensor with a channel width of 64, corresponding to the width of the Fourier layers. The structure of a single Fourier layer is illustrated in Fig.~\ref{model architecture illustration} and is mathematically expressed as: 
\begin{equation} 
    \label{fourier layer equation} 
    \mathcal{G} \brac{\mathcal{X}} = \sigma \brac{\mathcal{F}^{-1} \circ R \circ \mathcal{F} \brac{\mathcal{X}} + \mathcal{W}_{linear} \cdot \mathcal{X}}. 
\end{equation}
where $\mathcal{X}$ is the input tensor to the layer with a dimension of $\brac{\text{batch size}, \text{Fourier layer width}, H, W}$. In Eq.~\eqref{fourier layer equation}, $\mathcal{F}$ and $\mathcal{F}^{-1}$ represent the Fourier transform and its inverse, respectively. The learnable operator $R$ performs spectral filtering and weighting in the frequency domain, retaining 30 Fourier modes along each spatial dimension for this study. A residual connection, parameterised by linear weights $\mathcal{W}_{linear}$, introduces complementary learnable filters through a $1 \times 1$ 2D convolutional layer. The weighted residual is then summed with the inverse Fourier-transformed output of $R$ before being passed to the activation function $\sigma$. The final output layers, denoted as $\mathcal{Q}$, map the outputs from the Fourier layers to the target output space $\mathcal{U}$ of \model{FNO}.

\subsection{Physics-Informed Learning and Training Strategy}

\subsubsection{Physics-Informed Loss Construction}

A physics-informed loss $\mathcal{L}_{PI}$ has been introduced when training the models in this paper. For a dataset with $$\bigbrac{\alpha_i \brac{x,y}, \Phi_i \brac{x,y}}_{i=1}^{N}$$, where $\phi$ includes the flow field properties $\bigbrac{P, U_x, U_y}$ and the spatial domain is discretised with $n_x$ and $n_y$ points, the training loss is composed of a numerical error $\mathcal{L}_{data}$ and a physical error $\mathcal{L}_{physics}$ as shown in Eq.~\eqref{training loss equation}. The weighting coefficients $w_{1}$ and $w_{2}$ are the hyperparameters denoting the relative importance of two errors.
\begin{equation}
    \label{training loss equation}
    \mathcal{L}_{PI} = w_{1} \mathcal{L}_{data} + w_{2} \mathcal{L}_{physics} 
\end{equation}

The numerical error is defined as the mean of the mean squared error of the predictions (square of the $L^2$ norm) in the given dataset as shown by Eq.~\eqref{numerical loss definition}.
\begin{equation}
    \label{numerical loss definition}
    \mathcal{L}_{data} = \dfrac{1}{N} \sum_{i=1}^{N} \norm{\Phi_i - \widehat{\Phi}_i}_2^2 
\end{equation}
As shown by Eq.~\eqref{physics informed loss}, the physics-informed loss contains two terms, from the incompressible continuity and Navier Stokes equation, respectively. 
\begin{equation}
    \label{physics informed loss}
    \mathcal{L}_{physics} = \lambda_1 \mathcal{L}_{continuity} + \lambda_2 \mathcal{L}_{momentum} + \lambda_3 \mathcal{L}_{boundary}
\end{equation}
Parameters $\lambda$ are the scaling factors for each equation that ensure the numerical values of residuals are similar orders of magnitude.
The $\mathcal{L}_{continuity}$ is defined as the MSE of the residual of incompressible continuity equation (Eq.~\eqref{incompressible_ns}) as shown by Eq.~\eqref{continuity loss equation}.
\begin{equation}
    \label{continuity loss equation}
    \mathcal{L}_{continuity} = \dfrac{1}{N} \sum_{i=1}^{N} \norm{\divergence \widehat{\VecU}_i}_2 ^2
\end{equation}
Similarly, the residual from the steady incompressible Navier Stokes equation (Eq.~\eqref{ns equation} without the $\partialdiffs{}{t}$ term) is taken as part of the physics-informed loss as $\mathcal{L}_{momentum}$ in Eq.~\ref{momentum loss equation}
\begin{equation}
    \label{momentum loss equation}
    \mathcal{L}_{momentum} = \dfrac{1}{N} \sum_{i=1}^{N} \norm{\rho \widehat{\VecU}_i \cdot \nabla \widehat{\VecU}_i + \nabla \widehat{P}_i - \mu \laplace{\widehat{\VecU}_i} + \alpha_i \widehat{\VecU}_i}_2 ^2
\end{equation}

The $\mathcal{L}_{boundary}$ represents the boundary conditions, including the velocity inlet, pressure outlet and the no-slip wall boundary condition as shown by Fig.~\ref{fig:domain}.

All three model architectures (\model{AE}, \model{FNO}, and \model{UNET}) introduced in Section~\ref{Model Architecture method section} are trained with the physics-informed loss $\mathcal{L}_{PI}$ and will be denoted as \model{\dummy-PI}. In addition, another set of models is trained using the MSE loss $\mathcal{L}_{data}$ for comparison, which is denoted as \model{\dummy-data}.

\subsubsection{Dataset Generation and Preprocessing}
Each of the 100 individual topology optimization paths, defined by a set of 4 parameters in Table~\ref{tab:parameters}, generated 30 candidates. These candidates feature progressively advanced characteristics, ensuring a diverse dataset that includes both porous media flow with continuous transitions and internal channel configurations with abrupt changes between solid and fluid phases.

The resulting dataset contains a total of 3000 configurations, each with a resolution of $128 \times 128$. 
It is divided into training, validation, and testing subsets with a ratio of 8:1:1, respectively. 
The training dataset is used for supervised learning, while the validation dataset helps fine-tune the model's hyperparameters and prevent overfitting. 
The testing dataset, which remains completely unseen during training, is used to assess the performance of each model in Section~\ref{result section}.

\subsubsection{Training Configuration}

The models are trained for 200 epochs on an NVIDIA A100 (80 GB) GPU. The Adam optimizer \cite{kingma2014adam} was used with a learning rate of 0.001 for \model{FNO} and \model{UNET} and 0.0005 for \model{AE}, respectively. The parameters are $w_1$, $w_2$, $\lambda_1$ and $\lambda_2$ are tuned to minimise the validation dataset error. A batch size of 16 was used when training and each model takes approximately 20 min to train.

\section{Results}
\label{result section}

The Fourier Neural Operator proved suitable for learning non-periodic boundary conditions thanks to the linear bias term $\mathcal{W}_{linear}$ for Darcy Flow in \cite{li2020fourier}. 
In this paper, we extend this conclusion beyond Darcy flow in porous media by considering the Navier-Stokes-Brinkman model, which introduces significantly higher nonlinearities in the material and pressure/velocity fields.

The Fourier Neural Operator (FNO) is mesh-invariant because its Fourier layers operate directly in Fourier space, yielding consistent error across different resolutions.

\subsection{Model Performance}

\subsubsection{Model Error Comparisons}

\begin{table*}[width=\textwidth,cols=8,pos=ht]
    \setlength{\tabcolsep}{4pt}
    \renewcommand{\arraystretch}{1.5}
    \centering
    \caption{Model Error Comparison}

    \begin{tabular*}{\tblwidth}{@{} CCCCCCCC@{} } 
    
        \toprule

        \multirow{2}{*}{Channel Name}    &   \multirow{2}{*}{Metric} & \multicolumn{2}{c}{\model{AE}} & \multicolumn{2}{c}{\model{UNET}} & \multicolumn{2}{c}{\model{FNO}} \\
        
        & & $\mathcal{L}_{data}$ & $\mathcal{L}_{PI}$ & $\mathcal{L}_{data}$ & $\mathcal{L}_{PI}$ & $\mathcal{L}_{data}$ & $\mathcal{L}_{PI}$ \\ 
        
        \midrule
    
        
        $P$             & $R^2$                         & 0.9561719 & \textbf{0.9719298} & 0.8836706 & 0.9464990 & 0.9663823 & 0.9697812 \\
                        & MSE                           & 0.0412227 & 0.0264015 & 0.1094139 & 0.0503205 & 0.0316193 & \textbf{0.0284223} \\
                        & Relative $L^2$                & 0.2092571 & \textbf{0.1674784} & 0.3409388 & 0.2312260 & 0.1832587 & 0.1737473 \\
        
        $U_x$           & $R^2$                         & 0.9895563 & 0.9929830 & 0.9952475 & 0.9972704 & \textbf{0.9982153} & 0.9976720 \\
                        & MSE                           & 0.0103068 & 0.0069250 & 0.0046901 & 0.0026938 & \textbf{0.0017613} & 0.0022975 \\
                        & Relative $L^2$                & 0.1021510 & 0.0837348 & 0.0689079 & 0.0522208 & \textbf{0.0422318} & 0.0482332 \\
        
        $U_y$           & $R^2$                         & 0.9615440 & 0.9706253 & 0.9841518 & 0.9879774 & \textbf{0.9900161} & 0.9858974 \\
                        & MSE                           & 0.0416178 & 0.0317899 & 0.0171512 & 0.0130110 & \textbf{0.0108048} & 0.0152621 \\
                        & Relative $L^2$                & 0.1960247 & 0.1713276 & 0.1258608 & 0.1096209 & \textbf{0.0998976} & 0.1187151 \\
        
        Combined        & $R^2$                         & 0.9690546 & 0.9783670 & 0.9563944 & 0.9780651 & \textbf{0.9853207} & 0.9847239 \\
                        & MSE                           & 0.0310491 & 0.0217055 & 0.0437518 & 0.0220084 & \textbf{0.0147285} & 0.0153273 \\
                        & Relative $L^2$                & 0.1760318 & 0.1472244 & 0.2089476 & 0.1482472 & \textbf{0.1212900} & 0.1237444 \\
        
        PI Loss         & $\mathcal{L}_{continuity}$    & 0.0000454 & 0.0000005 & 0.0000075 & 0.0000008 & 0.0000059 & \textbf{0.0000004} \\
                        & $\mathcal{L}_{momentum}$      & 0.0150823 & 0.0006496 & 0.0084254 & 0.0014914 & 0.0044035 & \textbf{0.0002250} \\ 
                        
        \bottomrule
        
    \end{tabular*}

    \label{tab:errors}
\end{table*}            

Table~\ref{tab:errors} presents a comparison of model errors across different architectures and evaluation metrics. 
The models compared AE, U-Net, and FNO, each evaluated using two loss functions: MSE loss $\mathcal{L}_{data}$ as Eq.~\eqref{numerical loss definition} and physics-informed (PI) loss $\mathcal{L}_{PI}$ as Eq.~\eqref{physics informed loss}. 
The comparison is made across individual channels: $P$, $U_x$, and $U_y$, as well as for a combined loss across all channels and a separate physics loss, as defined in the previous section. Since all the cases are prescribed with the same boundary condition as shown in Fig.~\ref{fig:domain}, the boundary condition loss $\mathcal{L}_{boundary}$ are to the order of $10^{-10}$ and hence neglected in Table~\ref{tab:errors}.

When examining the relative $L^2$ error presented in Table~\ref{tab:errors}, we note the following characteristics:
\begin{itemize}[nolistsep]
    \setlength{\itemsep}{0pt}
    \setlength{\parskip}{0pt}
    
    \item FNO models consistently performs the best: Both \model{FNO-data} and \model{FNO-PI} exhibit the lowest losses across all channels ($P$, $U_x$, $U_y$, and Combined), indicating that they maintain superior accuracy compared to AE and U-Net models.
    
    \item PI loss improves performance for convolutional models: The PI model shows improvements between 12 and 32\% in loss over the data-only model for most channels, with the most significant improvement observed in the pressure ($P$) channel, where the physics-informed loss helps enforce continuity. The results will be further disucssed in Section~\ref{PI loss result section}
    
    \item Introduction of $\mathcal{L}_{PI}$ has no detrimental effect for spectral models: For the FNO models, the inclusion of PI loss results in a slight improvement of 5.19\% in the $P$ channel, but leads to an increase in loss by 12.94\% and 18.85\% for the $U$ channels.
    
    \item Physics-informed loss enforces domain consistency: The inclusion of physics-informed loss significantly reduces continuity loss by over 99\% and momentum loss by up to 95\% across all models, highlighting its effectiveness in enforcing physical laws like continuity and momentum consistency in predictions.
    
    \item \model{AE} outperforms \model{UNET} in certain cases: The AE’s simpler architecture with fewer parameters may lead to better generalization and reduced overfitting, especially when the dataset is small, less complex, or noise-free. The AE's reconstruction loss function aligns more closely with the task at hand.
    
    \item Model performance is physics-dependent: The \model{UNET} performs better on the $U_y$ channel, which contains more multiscale features that require both low- and high-level representations. Additionally, the gap between models narrows in other channels when the physics-informed loss is applied, as it effectively reduces the high continuity loss associated with the pressure channel.
    
\end{itemize}

In summary, FNO models (especially \model{FNO-PI} trained with PI loss) demonstrate the lowest Relative $L^2$ errors, reinforcing their advantage in accuracy over AE and U-Net models.

\subsubsection{Physics-Informed Error Comparison}
\label{PI loss result section}



\begin{landscape}
    \begin{figure}
        \centering
        \includegraphics[width=\linewidth]{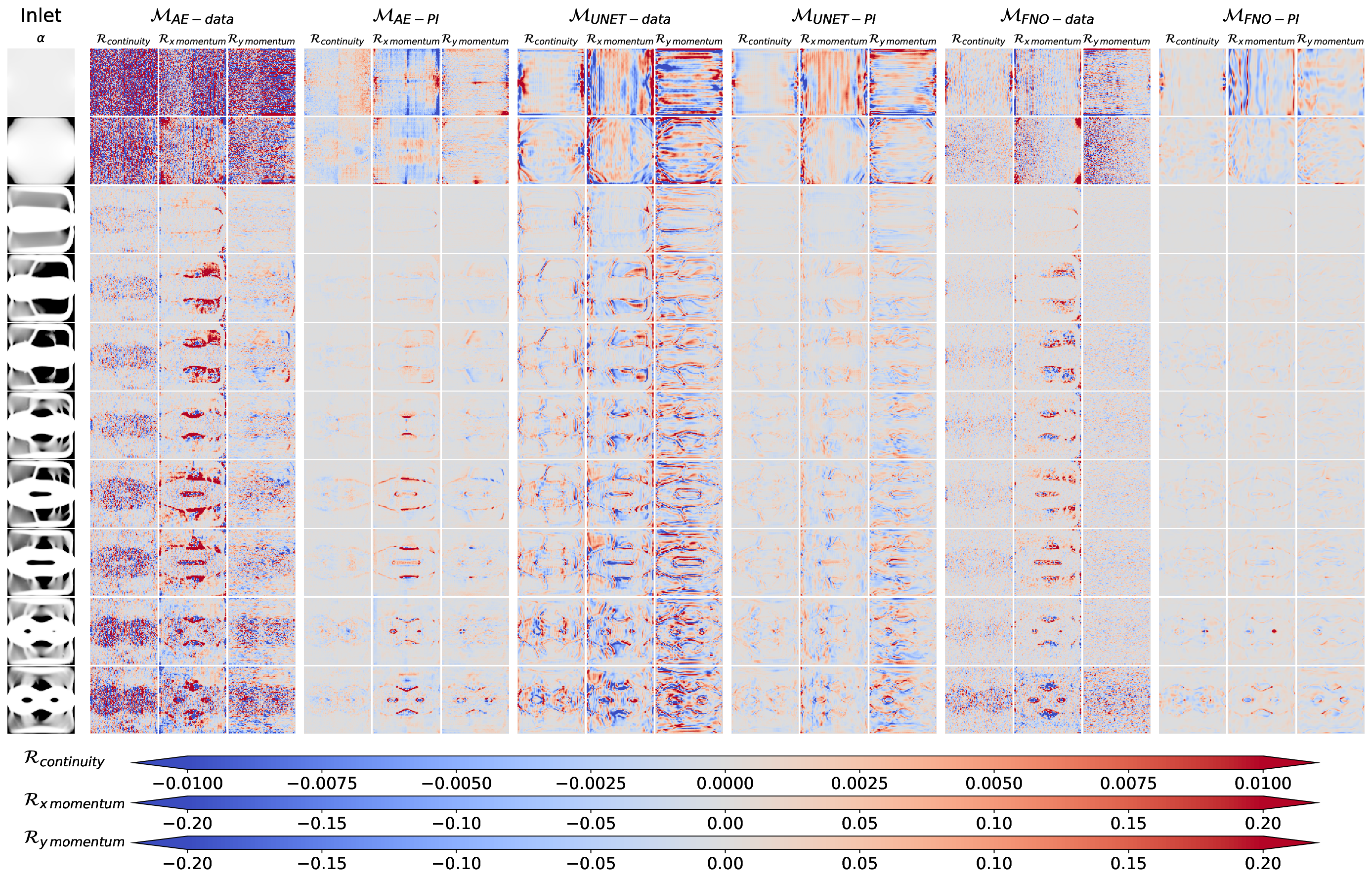}
        \caption{Physics residuals $\mathcal{R}_{continuity}$ and $\mathcal{R}_{momentum}$ for models that are trained with and without the physics-informed loss term $\mathcal{L}_{physics}$}
    \label{PI error improvement graph}
    \end{figure}
\end{landscape}

Figure~\ref{PI error improvement graph} illustrates the physics residuals, $\mathcal{L}_{continuity}$ and $\mathcal{L}_{momentum}$, across models trained with and without the physics-informed loss term, $\mathcal{L}_{physics}$. 
The inclusion of $\mathcal{L}_{physics}$ demonstrates a marked improvement in enforcing the governing equations, particularly evident in the reduced magnitude of residuals across the continuity and momentum terms as shown by Table~\ref{tab:errors}.

For all models, $\mathcal{L}_{physics}$ significantly mitigates noise and irregularities, especially in regions with complex flow behaviours. 
The AE and U-Net models, in particular, show considerable reductions in $\mathcal{L}_{continuity}$ when trained with $\mathcal{L}_{physics}$. 
This indicates enhanced alignment with mass conservation principles, which are critical for accurate flow prediction.

Moreover, the \model{FNO-PI} exhibits the smallest residuals overall, reflecting its inherent ability to capture global features with high fidelity. 
However, the inclusion of $\mathcal{L}_{physics}$ further improves the FNO's performance, particularly in challenging flow regimes where local complexity dominates. 
This improvement is most pronounced in the $\mathcal{L}_{momentum}$ terms, as shown by the sharp reduction in errors for both $x$ and $y$ momentum components.

In summary, the physics-informed loss term serves as a crucial enhancement across all architectures, ensuring compliance with physical laws and significantly improving the quality of predictions, particularly in complex and high-resolution flow scenarios. Therefore, the models that were trained using physics informed losses $\mathcal{L}_{PI}$ will be further studied in the rest of this paper for performance comparison.
A natural extension for future work is to explore architectures with inherent conservation properties, such as the Invariant Neural Operator (INO) \cite{liu2023invariant}, seeking a foundational machine learning model for convection in porous media.
This architecture ensures frame-invariant predictions through built-in kernel function invariance and aligns with Noether’s theorem to guarantee conservation of linear and angular momentum. 
This approach would further enhance compliance with physical principles, particularly in systems with heterogeneous properties or unknown governing equations.

\begin{figure*}[pos=h]
    \centering
    \includegraphics[width=\textwidth]{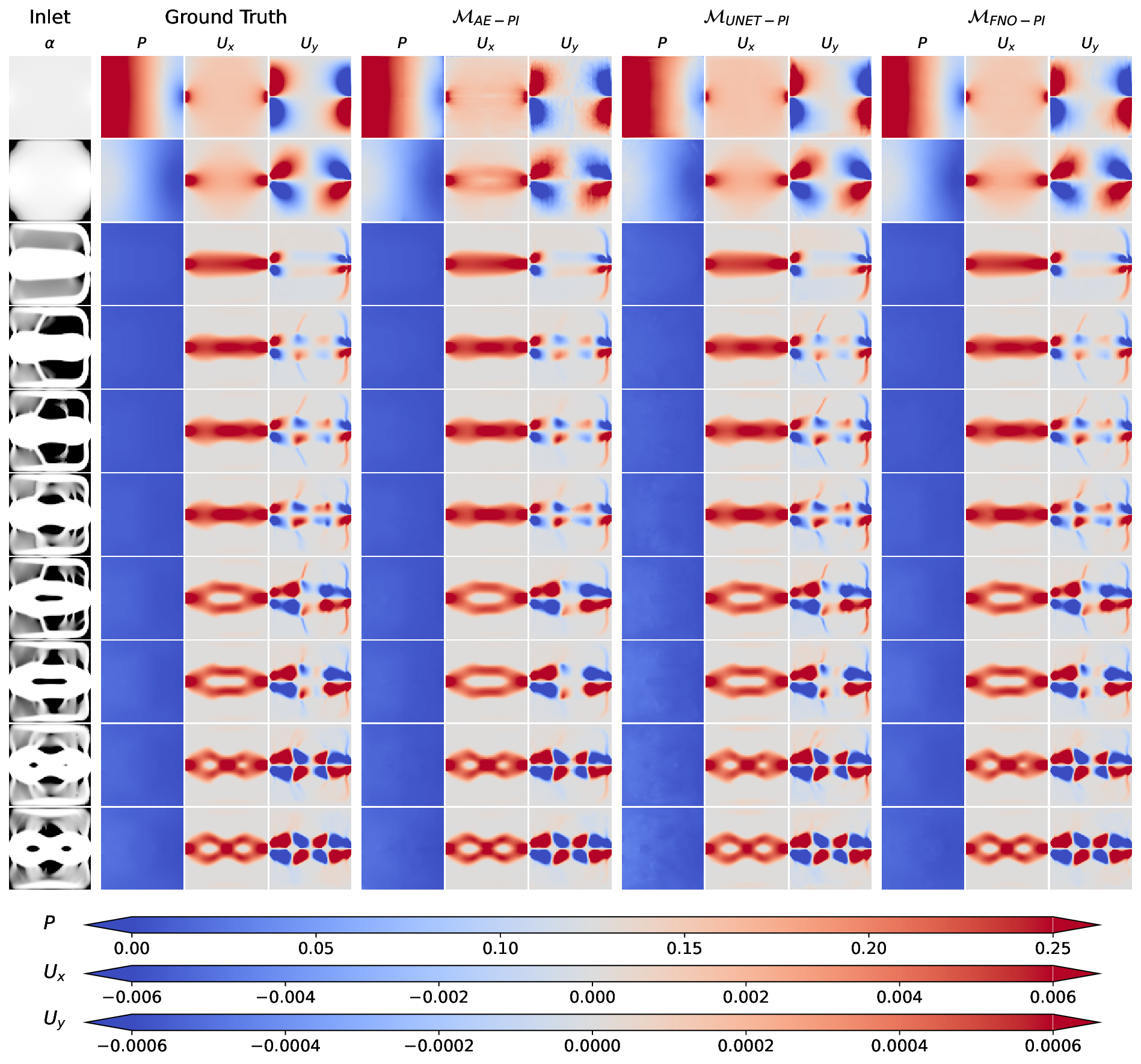}
    \caption{Comparison of predictions for different material field inputs across the $P$, $U_x$, and $U_y$ channels using three physics-informed machine learning architectures: autoencoder (AE), U-Net, and Fourier Neural Operator (FNO).}
    \label{pred graph}
\end{figure*}

\begin{figure*}
    \centering
    \includegraphics[width=\textwidth]{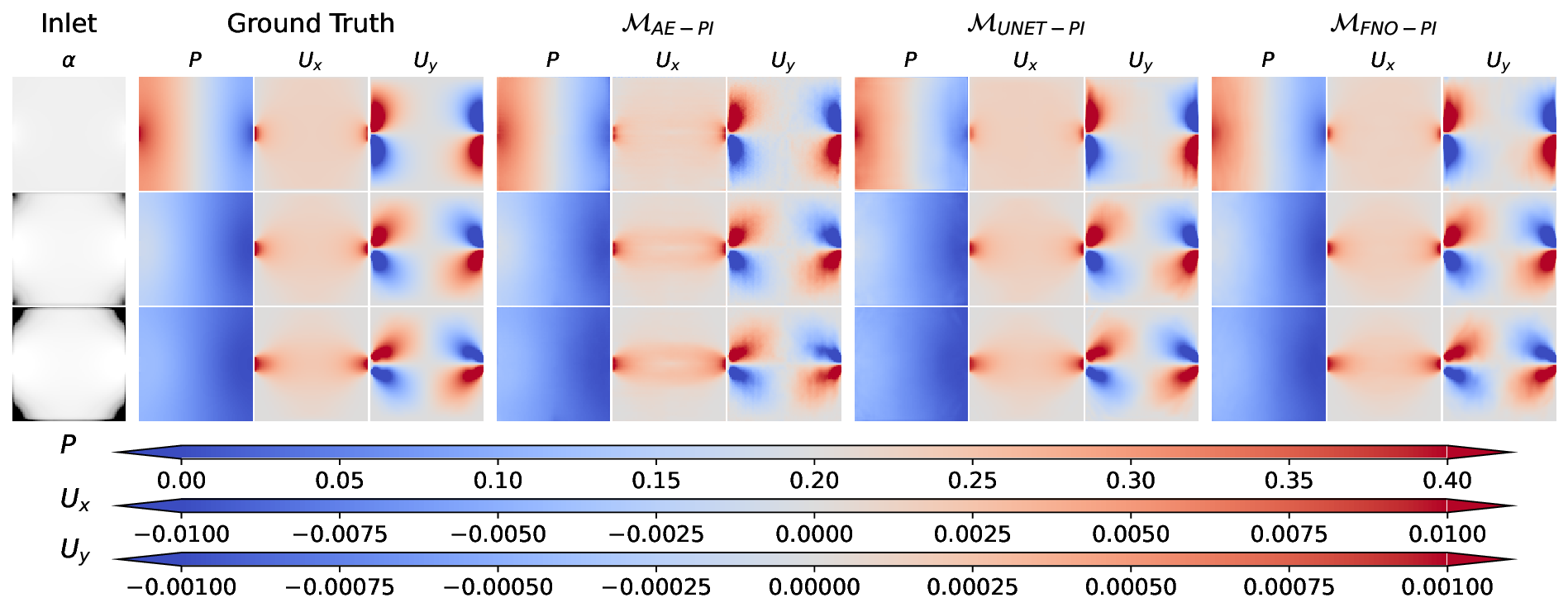}
    \caption{Uniform Topologies: most homogeneous configurations, with minimal variation in material distribution}
    \label{fig:ut}
\end{figure*}

\begin{figure*}
    \centering
    \includegraphics[width=\textwidth]{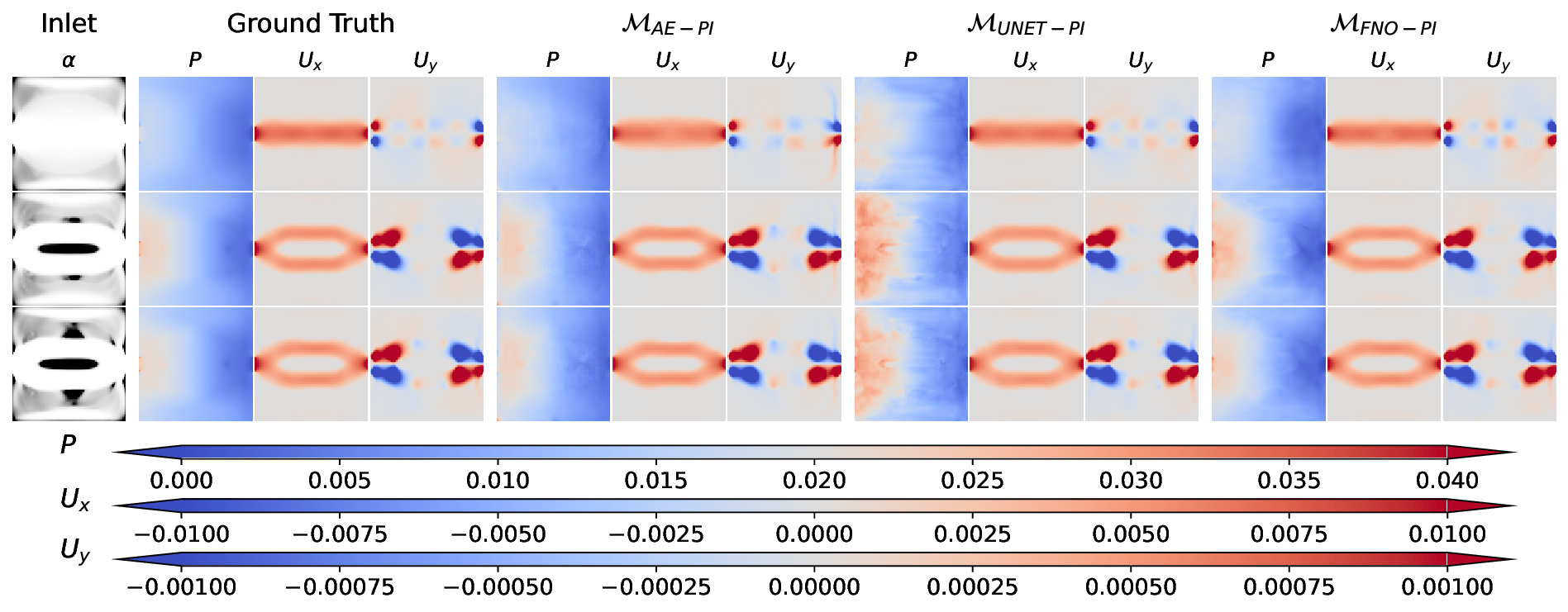}
    \caption{Gradual Material Variations: topologies with minor deviations and linear transitions between material properties emerging in the flow structure}
    \label{fig:gmv}
\end{figure*}

\begin{figure*}
    \centering
    \includegraphics[width=\textwidth]{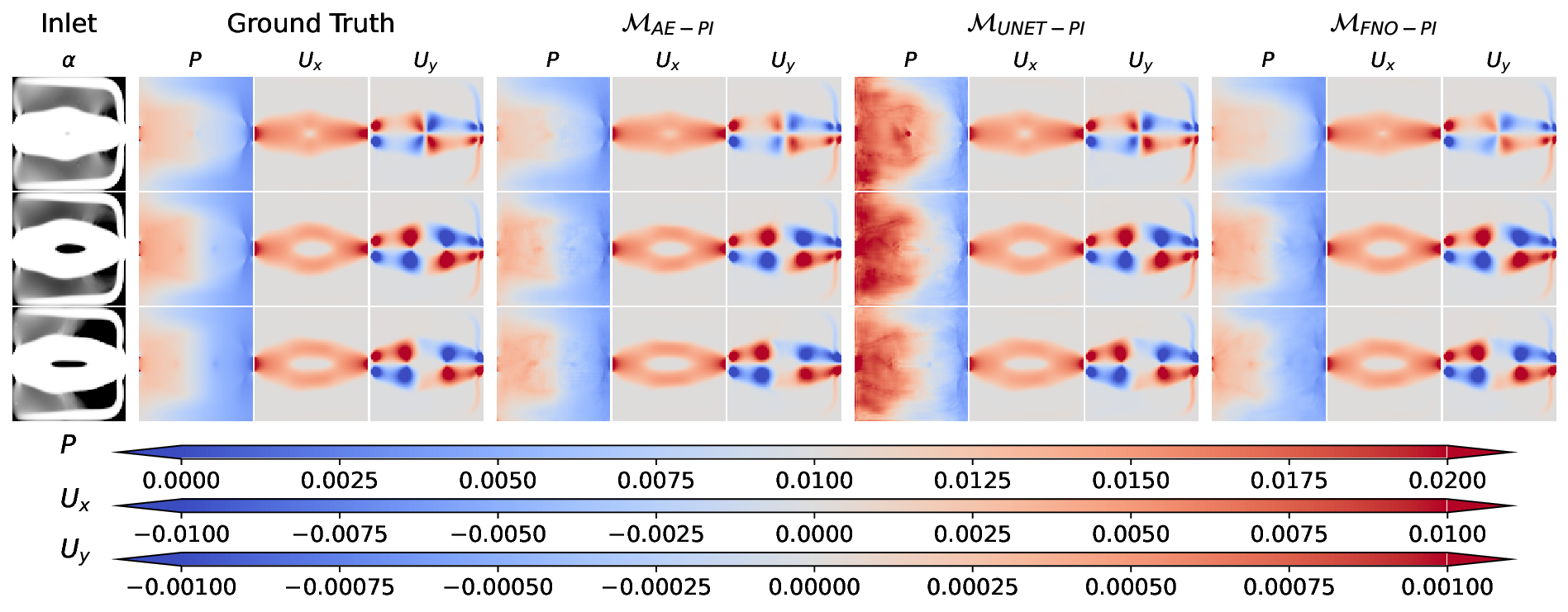}
    \caption{Developing Channels: configurations with partially formed channels, representing significant directional flow pathways}
    \label{fig:dc}
\end{figure*}

\begin{figure*}
    \centering
    \includegraphics[width=\textwidth]{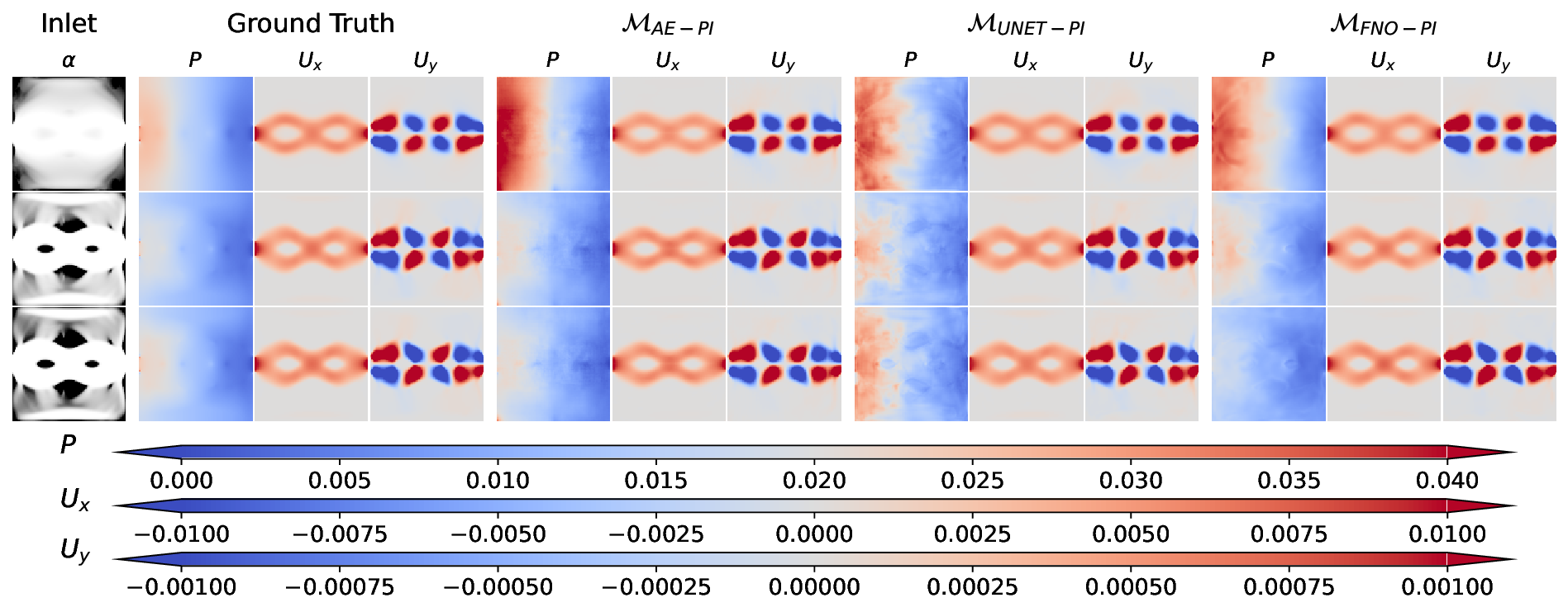}
    \caption{Clustered Variations: topologies where heterogeneities start forming distinct clusters, introducing localized complexity}
    \label{fig:cv}
\end{figure*}

\begin{figure*}
    \centering
    \includegraphics[width=\textwidth]{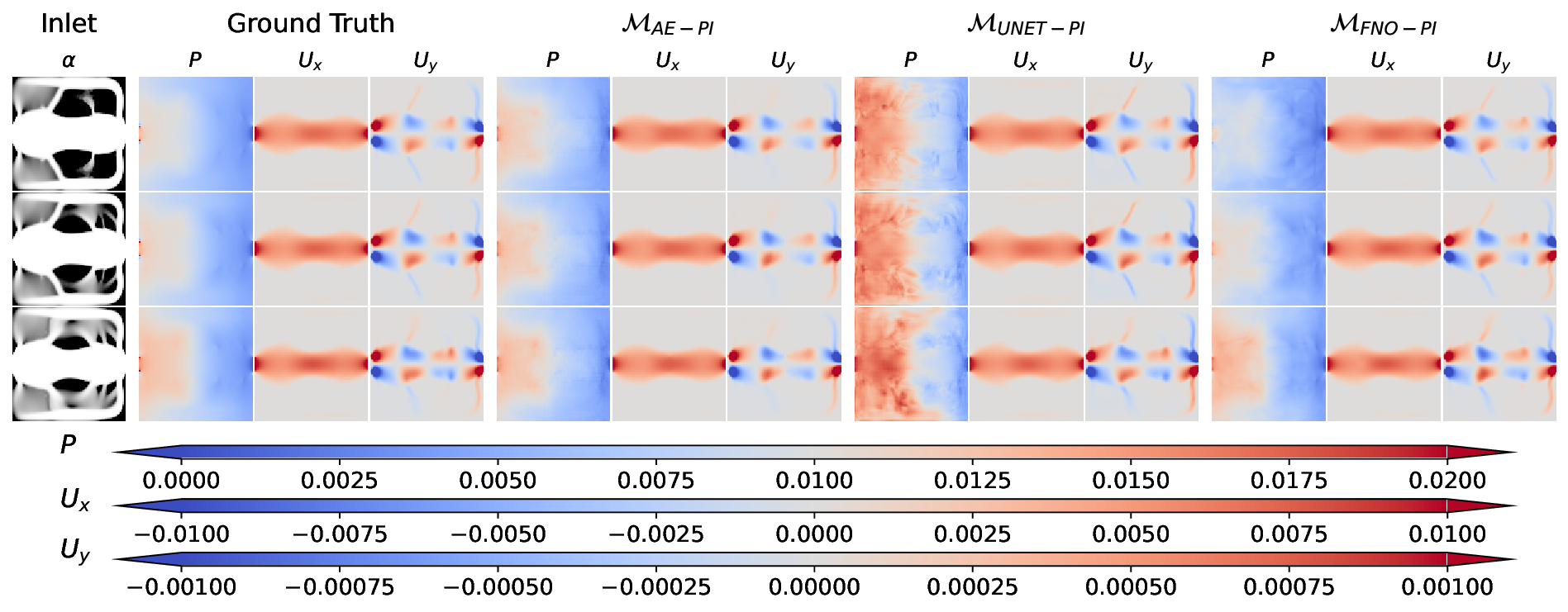}
    \caption{Mature Channel Networks: most evolved topologies, characterized by well-defined and interconnected channel structures}
    \label{fig:mcn}
\end{figure*}

\begin{figure*}[pos=h!]
    \centering
    
    \begin{subfigure}{0.85\linewidth}
        \centering
        \includegraphics[width=\textwidth]{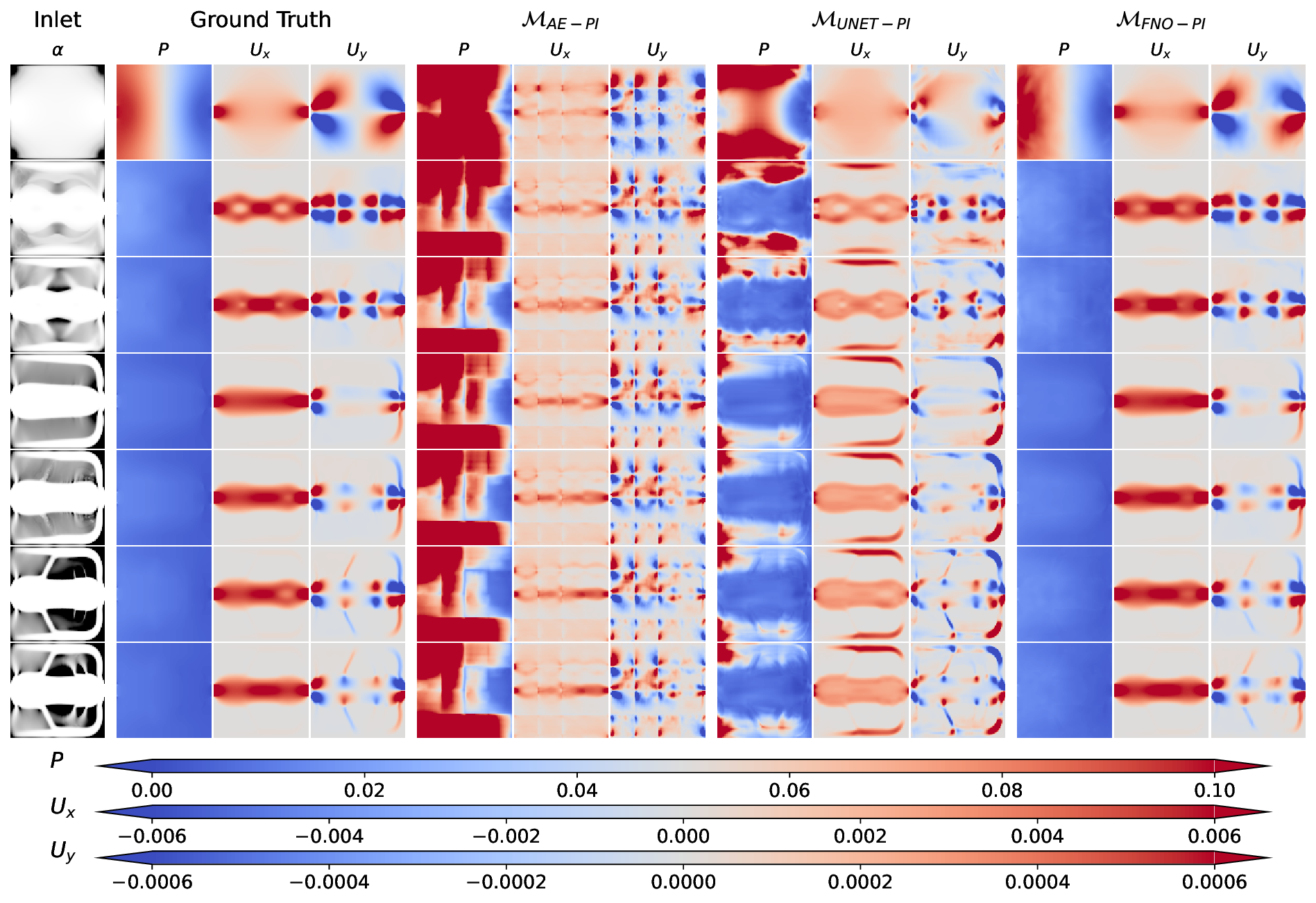}
        \subcaption{Results with resolution $256 \times 256$}
        \label{pred 256 graph}
    \end{subfigure} \\
    
    \begin{subfigure}{0.85\linewidth}
        \centering
        \includegraphics[width=\textwidth]{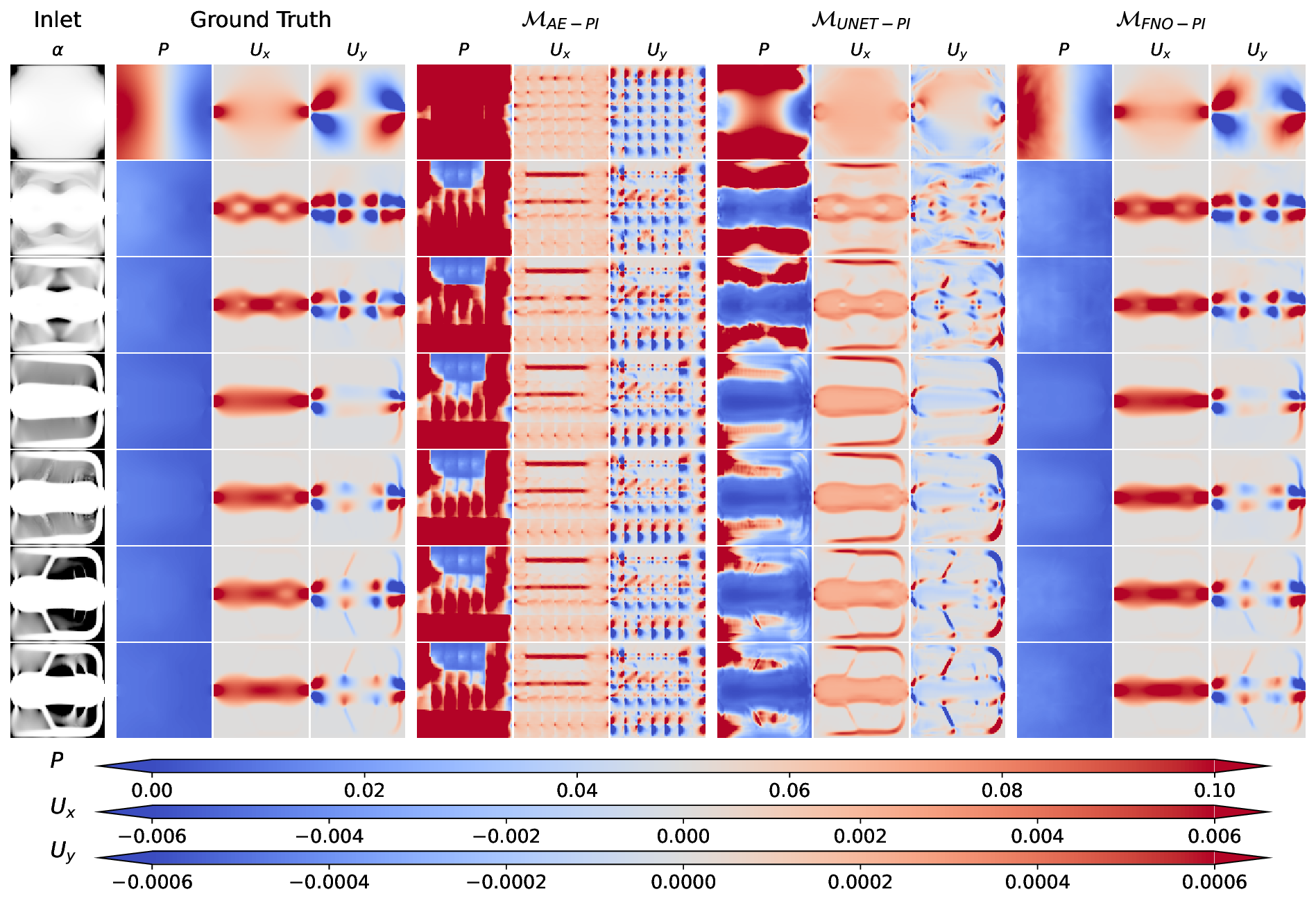}
        \subcaption{Results with resolution $400 \times 400$}
        \label{pred 400 graph}
    \end{subfigure} 
    
    \caption{Model predictions of \model{AE-PI}, \model{UNET-PI}, and \model{FNO-PI} for a testing dataset with different resolutions. }
    \label{mesh invariant graph}
    
\end{figure*}

\subsubsection{Model Predictions}

Figure~\ref{pred graph} complements the error comparisons in Table~\ref{tab:errors}, offering insights into how the models capture spatial features and enforce physical consistency under different inputs. The trends observed in the predictions align with the quantitative error metrics shown in Table~\ref{tab:errors}:
\begin{itemize}[nolistsep]
    \setlength{\itemsep}{0pt}
    \setlength{\parskip}{0pt}
    
    \item The FNO models consistently produce results that closely align with the ground truth across all channels. Their Fourier-based architecture excels at capturing both complex and quasi-linear patterns, demonstrating robustness across inputs with varying levels of evolutionary complexity.
    
    \item While AE models perform well in scenarios with simpler spatial features, their predictions degrade for more complex multiscale patterns seen in the $U_y$ channel.
    
    \item U-Net predictions are particularly strong in the $U_y$ channel due to its ability to handle multiscale complexity, but the $P$ channel exhibits strong overfitting tendencies for more complex material features depicted in the bottom two rows of the figure.
    
\end{itemize}



To further understand the models performance, the performance of the models is assessed across various flow regimes, which correspond to distinct categories of material property configurations within the dataset. 
These configurations represent a progression from uniform topologies to highly evolved channel networks, reflecting increasing complexity in material heterogeneity and flow physics.

The following categories capture the diversity of configurations present in the dataset:
\begin{itemize}[nolistsep]
    \setlength{\itemsep}{0pt}
    \setlength{\parskip}{0pt}
    
    \item Uniform Topologies: These represent the most homogeneous configurations, with minimal variation in material distribution (Fig.~\ref{fig:ut}).
    
    \item Gradual Material Variations: Topologies with minor deviations and linear transitions between material properties, introducing gradual changes in the flow structure (Fig.~\ref{fig:gmv}).
    
    \item Clustered Variations: Configurations where heterogeneities form distinct clusters, creating localized complexities in flow dynamics (Fig.~\ref{fig:cv}).
    
    \item Developing Channels: Representations of partially formed channels, indicating significant directional flow pathways and intermediate material evolution (Fig.~\ref{fig:dc}).
    
    \item Mature Channel Networks: The most evolved and intricate topologies, featuring well-defined, interconnected channel structures that dominate flow behavior (Fig.~\ref{fig:mcn})
    
\end{itemize}

In the figure depicting Uniform Topologies (Fig.~\ref{fig:ut}), the \model{AE} performs poorly, especially in the $U_x$ channel. 
This is likely due to AE's difficulty in capturing the abrupt transitions between the inlet/outlet regions and the rest of the field, which are key features of the uniform material fields that do not have flow guiding (\textit{nozzle}) features.
On the other hand, both the \model{UNET} and \model{FNO} show similar performance, suggesting that their architectures are better suited to handling such transitions, as they can more effectively capture the non-uniformities in the velocity field.

As shown in Fig.~\ref{fig:gmv}, all models exhibit difficulty in accurately predicting the parabolic pressure profiles, resulting in noticeable artifacts. 
While the \model{AE}'s performance improves relative to the previous case, benefiting from the presence of more guiding features for generating uniform velocity fields, it still underperforms compared to the \model{UNET} and \model{FNO}. 
Notably, \model{UNet} struggles particularly with the pressure prediction due to overfitting, introducing noise that distort the flow structure.

As shown in Fig.~\ref{fig:dc}, the flow pathways become more defined, and pressure variations are less abrupt, allowing for clearer directional flow. 
In this case, the \model{AE} outperforms the \model{UNET}, which continues to struggle with overfitting the pressure profile. 
The \model{AE} benefits from the improved structural guidance provided by the developing channels, leading to simpler flow features and better prediction accuracy, particularly in the pressure field.
However, the \model{FNO} remains the most accurate model for this class of configurations.

As shown in Fig.~\ref{fig:cv}, the topologies with clustered variations introduce more localized complexity, particularly in the pressure field predictions. 
This results in all models, including \model{FNO}, struggling with accurate pressure predictions. 
However, all models still provide reliable velocity predictions, which are arguably more relevant for accurate convection calculations, which play a crucial role in thermal design applications.

As shown in Fig.~\ref{fig:mcn}, in the mature channel networks, characterized by well-defined and interconnected channel structures, the AE performs well but struggles with finer variations in the pressure field, providing a relatively stiff response. 
The \model{UNet} continues to overfit the pressure profile, a common issue for profiles with smaller variations. 
The \model{FNO} demonstrates steady performance for this class, recovering from the earlier drop in prediction accuracy seen in the previous figure. 
This improvement can be attributed to the now more gradual pressure profiles driven by high-resolution channel features, allowing the model to make use of its spectral nature to capture complex patterns with greater accuracy, particularly in regions with subtle periodic variations in both pressure and velocity.

\subsection{Mesh Independence of Models}

For traditional neural networks, such as \model{AE} and \model{UNET}, the mesh resolution directly impacts both the input and output sizes of the model, which are tied to fixed spatial grids. 
These networks operate on a finite-dimensional input-output space, specifically:
\begin{equation} \mathcal{A} \subset \mathbb{R}^{1 \times H \times W} ~~,~~ \mathcal{U} \subset \mathbb{R}^{3 \times H \times W} \end{equation}
where $H$ and $W$ are the height and width of the grid, with both being 128 in the training data. 
As a result, changes in the spatial resolution of the input (e.g. $256 \times 256$ and $400 \times 400$) will require resizing the input and output data, and the model performance is highly sensitive to the resolution change. 
This can lead to performance degradation or require retraining when the resolution differs from the training data.
This sensitivity to mesh resolution is a crucial consideration for design purposes, as conventional topology optimization using finite element (FE) methods is notoriously mesh-dependent, often requiring fine-tuned meshes for reliable results.

In contrast, the \model{FNO}, being a neural operator, works in function spaces, where both the input and output domains are defined on continuous function spaces over the domain $\Omega$. Specifically, the input and output spaces are described as:
\begin{equation} \mathcal{A} \subset L^2\brac{\Omega, \mathbb{R}} ~~,~~ \mathcal{U} \subset L^2\brac{\Omega, \mathbb{R}^3} \end{equation}

As a result, \model{FNO} is inherently mesh-invariant, as it can handle changes in resolution without requiring retraining. 
This property is a key advantage of neural operators, as they generalize to different grid sizes without any performance degradation, as demonstrated in Fig.~\ref{mesh invariant graph}. 
The mesh independence is a direct consequence of the continuous nature of neural operators, which allow them to process data represented on different spatial grids without being tied to a specific discretization.

In Fig.~\ref{mesh invariant graph}, we show the model predictions for \model{AE}, \model{UNET}, and \model{FNO} on testing datasets with resolutions of $256 \times 256$ and $400 \times 400$. 
The predictions of \model{AE} and \model{UNET} exhibit noticeable differences with varying resolutions, as expected, due to their dependence on fixed input sizes. 
Between the two neural network models, \model{UNET} outerperforms \model{AE} on grid-independence due to its use of skip connections, which allow the network to retain more detailed spatial information.
On the other hand, \model{FNO} maintains consistent performance across the two resolutions, further validating its mesh-invariant property.
This property is particularly advantageous for thermal design, as it eliminates the need for mesh refinement typically required in conventional finite element methods.





\subsection{Prediction Speed}

In this section, we evaluate the computational efficiency of each model by comparing the average calculation time for one design topology across different resolutions. 
The results in Table~\ref{tab:time} present the time in seconds for each method. 

The machine learning models (\model{AE-PI}, \model{UNET-PI}, and \model{FNO-PI}) are assessed in both CPU and GPU modes to highlight the performance differences across hardware configurations. The finite element method solver implemented in FEniCS is executed on a single CPU for baseline comparison.
The \model{FNO-PI} model demonstrates a significant speedup over the FEniCS solver, with improvements ranging from approximately \textbf{100 times faster} at resolution $128 \times 128$ to about \textbf{1000 times faster} at resolution $400 \times 400$ when running on the GPU. 
This highlights the substantial efficiency gains that machine learning models can offer over conventional methods like FEniCS, particularly when leveraging GPU acceleration.

As shown in Table~\ref{tab:time}, the baseline run for all models is performed on the GPU. 
All three models operate within the same order of magnitude in this regime, with substantial speedups observed over the CPU times.
For \model{AE-PI} and \model{UNET-PI}, switching to the CPU results in a noticeable increase in computation time. For example, at a resolution of $128 \times 128$, the GPU time for \model{MAE-PI} is 0.0010 seconds, while the CPU time is 0.0142 seconds, which is roughly 14 times slower on the CPU. Similarly, \model{UNET-PI} shows about a 14-fold increase in computation time on the CPU compared to the GPU.

In contrast, \model{FNO-PI} demonstrates a much more dramatic improvement when moving from CPU to GPU. 
At a resolution of $128 \times 128$, the GPU time for \model{FNO-PI} is 0.0020 seconds, while the CPU time is 0.1089 seconds—roughly 55 times slower on the CPU. 
While not the fastest model in absolute terms, \model{FNO-PI} benefits significantly from GPU acceleration, with a substantial speedup compared to its CPU performance. 
This indicates that \model{FNO-PI} is relatively efficient and can make the most of GPU resources, making an excellent trade-off between accuracy and speed.

\begin{table*}[width=.9\textwidth,cols=9,pos=ht]
    \setlength{\tabcolsep}{6pt}
    \renewcommand{\arraystretch}{1.5}
    \centering
    \caption{Comparison of average calculation speed for one design topology in seconds}
    \label{tab:time}
    \begin{tabular*}{\tblwidth}{@{} CCCCCCCCC@{} }
        \toprule
        \multirow{2}{*}{Resolution} & \multirow{2}{*}{FEniCS Solver} & \multicolumn{2}{c}{\model{AE-PI}} & \multicolumn{2}{c}{\model{UNET-PI}} & \multicolumn{2}{c}{\model{FNO-PI}} \\
                                    &                                & CPU        & GPU       & CPU         & GPU        & CPU        & GPU        \\ 
        \midrule
        
        128 & 17.8406 & 0.0142 & 0.0010 & 0.0343 & 0.0025 & 0.1089 & 0.0020 \\
        256 & 82.0864 & 0.0366 & 0.0011 & 0.1400 & 0.0028 & 0.1338 & 0.0021 \\
        400 & 212.1420 & 0.1057 & 0.0010 & 0.3125 & 0.0024 & 0.2292 & 0.0021 \\ 
        \bottomrule
    \end{tabular*}
\end{table*}

\begin{figure}
    \centering
    \includegraphics[width=0.7\linewidth]{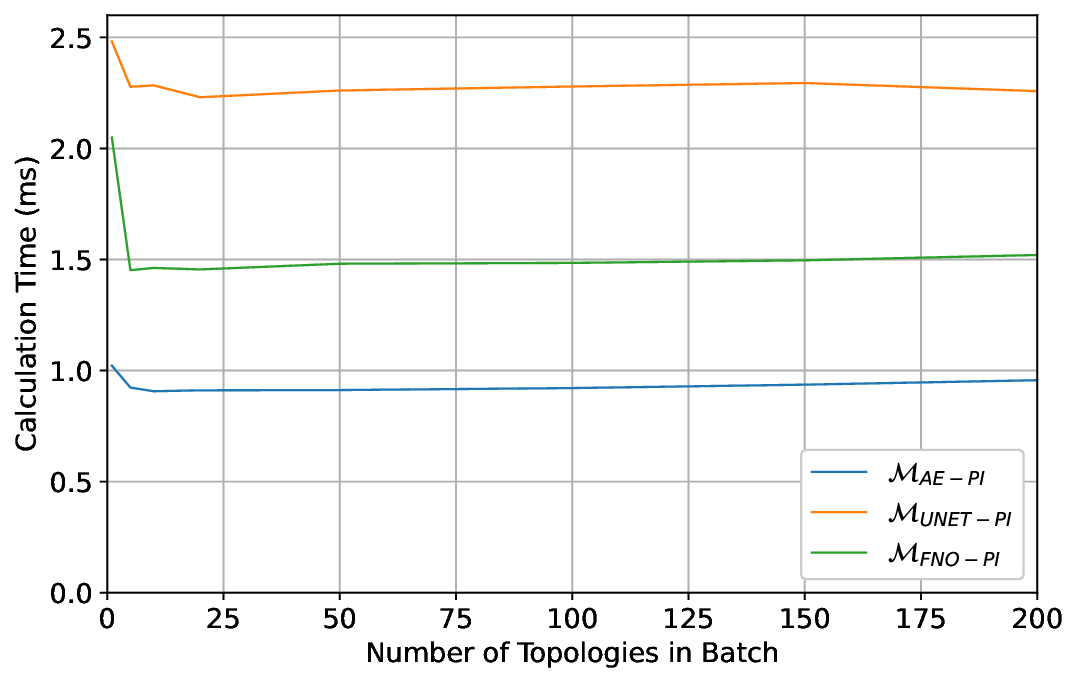}
    \caption{Calculation speed comparison between models on GPU}
    \label{fig:batch speed}
\end{figure}

As illustrated in Fig.~\ref{fig:batch speed}, the computational time of the models on the GPU varies with the number of topologies in the batch. 
While increasing the batch size generally results in more efficient GPU utilization and faster computation times, small batches tend to show worse performance. 
This is because GPUs are designed to handle large amounts of data in parallel, and when batch sizes are small, only a small fraction of the GPU’s cores are active. 
Consequently, the GPU’s parallel processing capabilities are underutilized, leading to a less efficient execution for small batches.

Interestingly, the line in the figure remains almost horizontal as the batch size increases, indicating that the flow-solving process in machine learning models can be highly parallelized. 
Unlike conventional solvers like finite difference or finite element methods, which typically solve each configuration sequentially, machine learning enables batched evaluation, allowing multiple topologies to be processed simultaneously.
This results in the computational load being effectively distributed across the GPU’s cores, enabling an exponential reduction in computation time as the number of topologies increases.

\section{Conclusion}

In this paper, we applied state-of-the-art machine learning models to predict porous flow behavior, demonstrating their effectiveness as a proof of concept for solving complex flow problems in porous media. 
Our work focused on evaluating the performance of three machine learning architectures -- Autoencoder (\model{AE}), U-Net (\model{UNET}), and Fourier Neural Operator (\model{FNO}) -- in the context of fluid flow prediction in porous media.
This study goes beyond traditional Darcy-based approaches by capturing more complex flow behaviors that are often encountered in heterogeneous and designed porous media using the Navier-Stokes-Brinkman model.

We validated and compared the models across various flow regimes, showing that the FNO outperforms both the AE and U-Net in terms of accuracy. 
The $R^2$ values for the FNO were consistently high across all channels, with the best results found in the $U_x$ channel, achieving an $R^2$ of 0.998, compared to 0.989 for \model{AE-data} and 0.995 for \model{UNET-data}. 
Additionally, the MSE (mean-squared error) values for the FNO ranged from 0.0017 to 0.042, significantly smaller than those of the next best models in the literature for low-Re flows.
These findings demonstrate the FNO’s superior ability to capture complex flow patterns in porous media.

Additionally, we highlighted the mesh-invariant nature of the FNO. 
Unlike traditional methods (both machine learning and physics solvers), which are sensitive to changes in mesh resolution, the FNO demonstrated consistent performance across different resolutions without the need for retraining. 
This mesh independence makes the FNO particularly useful in practical applications, where varying grid sizes are often required.

A significant finding of this study was the computational efficiency of the machine learning models, particularly when leveraging GPU acceleration. 
The FNO achieved speedups of up to 1000 times compared to finite element solvers like FEniCS, especially at higher spatial resolutions. 
The ability to perform batched evaluations further improves computational efficiency, allowing for the parallel processing of multiple topologies and significantly reducing overall computation time in the context of thermal design.

In conclusion, this work demonstrates that machine learning models, particularly the FNO, offer a promising alternative to traditional methods for predicting porous flow behavior. 
The models are not only faster and more efficient but also scalable and mesh-invariant, making them highly suitable for real-world thermal design applications. 
Moreover, the physics-informed approach ensures that the predictions remain consistent with the underlying physical principles, further enhancing their reliability. 
This opens the door for future applications of machine learning in computational fluid dynamics and multi-physics problems, where the methodology can be extended to cover more complex scenarios.









\section{Declaration of competing interest}
The authors declare that they have no known competing financial interests or personal relationships that could have appeared to influence the work reported in this paper.

\section{Data availability}
The data that support the findings of this study are available from the corresponding author upon reasonable request. The code used for this paper can be found in the GitHub repository: \url{https://github.com/milowangjinhong/AI_Predictor_for_Flow_Through_Porous_Media} .

\section{Acknowledgement}
The authors would like to thank Professor Sibo Cheng at Institut Polytechnique de Paris for his insights and suggestions in improving this paper.

\bibliographystyle{cas-model2-names}

\bibliography{cas-refs}



\end{document}